# W-Strings from Affine Lie Algebras

José M. Figueroa-O'Farrill [†]

*Department of Physics, Queen Mary and Westfield College*
*Mile End Road, London E1 4NS, UK*

## ABSTRACT

The most disappointing aspect of W-strings is probably the fact that at least for the known models one does not recover a physical spectrum that differs much from that of ordinary string theory. It is hoped that this is not an intrinsic shortcoming of W-string theory, but rather that it is a consequence of the W-algebra realizations that have been chosen. In this note we point out a whole new class of possible W-strings built from representations of affine Lie algebras via (quantum) Drinfel'd-Sokolov reduction. Explicitly, we construct a BRST operator in terms of $\widehat{sl}_3$ currents which computes the physical spectrum of a $\mathsf{W}_3$-string. As a special case of this construction, if we take a free-field realization of $\widehat{sl}_3$, we recover the 2-scalar $\mathsf{W}_3$-string. These results generalize to any W-algebra which can be obtained via quantum Drinfel'd-Sokolov reduction from an affine algebra.

---

[†] e-mail: jmf@strings1.ph.qmw.ac.uk

## Introduction

Soon after the BRST approach to string theory caught on, it was realized that there was nothing sacred about the 26 free scalar fields in the usual $\sigma$-model background for the bosonic string. One simply takes any conformal field theory with $c=26$, and defines the physical states of that "string" theory, as the BRST cohomology of the Virasoro algebra with values in the representations appearing in this CFT. The earlier steps in this approach were rather conservative: a few of the bosons were kept in order to maintain some semblance of a string theory, and the rest was taken to be any CFT of the correct central charge. But it did not take long for bolder proposals to emerge in which the bosons were no longer deemed necessary. This led to the notion of "strings without strings." Similar considerations were applied to superstring theories, which in this context simply corresponds to the computation of the BRST cohomology of another choice of algebra: a superconformal algebra this time. With the construction of the BRST charge for the $\mathsf{W}_3$-algebra [1], it was realized that one could now study intriguing generalizations of string theory, in which the (super)Virasoro algebra is substituted for a W-algebra. These generalized string theories are known as W-strings and have recently been reviewed in [2].

Contrary to the Virasoro or superVirasoro algebras, for which a general theory of BRST cohomology exists (as does for most Lie algebras), for W-algebras our body of knowledge consists of a few isolated examples and perhaps a few results of a more general nature. Since there is no general result which guarantees even the existence of the BRST operator, one is forced to tackle the problem case by case. At this point in time, the BRST operator is known to exist for only a few algebras: $\mathsf{W}_3$ [1], $\mathsf{W}_4$ [3] [4], $\mathsf{W}(2,4)$ [4] [5], $\mathsf{W}(2,6)$ [5] and some quadratically nonlinear algebras [6]—all these results having been obtained by explicit construction, which explains why only the smallest W-algebras have been studied until now.

Of these W-string theories, it is the $\mathsf{W}_3$-string that has received the most attention. Because many of the techniques for the computation of the BRST cohomology of Lie algebras do not exist for the case of $\mathsf{W}_3$ (although see [7] and [8]), one is forced to consider explicit CFT realizations of $\mathsf{W}_3$. The best known realizations are those constructed by Romans [9] in terms of a Virasoro algebra and one free boson, generalizing the 2-boson realization of Fateev and Zamolodchikov [10]. After a long series of attempts by a number of people (see [2] for a list of references) the $\mathsf{W}_3$ BRST cohomology of the Romans realizations has finally been computed [11]. The result, which had already been foreseen, is somewhat disappointing: the $\mathsf{W}_3$ string seems to be essentially a noncritical bosonic string coupled to the Ising model. There have been other attempts at constructing different realizations of $\mathsf{W}_3$ [12], but the ensuing W-string spectrum is again essentially that of an ordinary bosonic string or two.

One would hope that this is not an intrinsic shortcoming of the $W_3$-string, but rather a property of the realizations that have been hereto studied. It therefore behooves us to find new realizations of $W_3$.

One way to do this is to use quantum Drinfel'd-Sokolov reduction [13] [14] [15] [16]. Classically, the $W_3$-algebra can be constructed (or in fact, defined) as the fundamental Poisson brackets of a Poisson manifold which arises as the hamiltonian reduction of the dual of the affine Lie algebra $\widehat{sl}_3$, equipped with the natural Kirillov bracket. In its quantum setting, this hamiltonian reduction becomes BRST cohomology. Indeed, the $W_3$-algebra can be recovered as the cohomology of a BRST operator associated to a maximal nilpotent subalgebra of $\widehat{sl}_3$ and a particular 1-dimensional representation (that is, a character) of this subalgebra [17] [15]. Naturally, this procedure extends to representations: sending representations of $\widehat{sl}_3$ to representations of $W_3$. In particular, as shown in [17], the standard free field realization of $\widehat{sl}_3$ gets sent to the 2-scalar realization of $W_3$. It is not known whether any other one of the Romans realizations can be obtained in this way; not that this is very relevant for our present purpose, since what we are after are precisely other realizations with more interesting $W_3$-string spectrum.

In this note we use these ideas to construct a BRST operator in terms of $\widehat{sl}_3$ currents such that its cohomology computes the physical spectrum of a $W_3$-string.

The BRST complex for the $W_3$-algebra

Let us start by briefly reviewing the construction of the BRST charge for Zamolodchikov's $W_3$ algebra:

$$
\begin{aligned}
T(z)T(w) &= \frac{c_m/2}{(z-w)^4} + \frac{2T(w)}{(z-w)^2} + \frac{\partial T(w)}{z-w} + \text{reg.} \\
T(z)W(w) &= \frac{3W(w)}{(z-w)^2} + \frac{\partial W(w)}{z-w} + \text{reg.} \\
W(z)W(w) &= \frac{c_m/3}{(z-w)^6} + \frac{2T(w)}{(z-w)^4} + \frac{\partial T(w)}{(z-w)^3} \\
&+ \frac{1}{(z-w)^2}[2b\Lambda(w) + \tfrac{3}{10}\partial^2 T(w)] \\
&+ \frac{1}{z-w}[b\partial\Lambda(w) + \tfrac{1}{15}\partial^3 T(w)] + \text{reg.}
\end{aligned}
\quad (1)
$$

where the parameter $b$ is related to the central charge $c_m$ by $b = 16/(22+5c_m)$, and $\Lambda = (TT) - \tfrac{3}{10}\partial^2 T$. We follow the "point-splitting" definition of normal-ordering [18]

$$(AB)(z) = \oint_{C_z} \frac{dx}{2\pi i} \frac{A(x)B(z)}{x-z} \ . \quad (2)$$

Normal-ordering is of course not associative, and in our conventions $(ABC) \equiv (A(BC))$ and sometimes we even drop the parenthesis if there is no source of confusion.

We can expand $T(z) = \sum_n L_n z^{-n-2}$ and $W(z) = \sum_n W_n z^{-n-3}$ and the algebra of modes is easily worked out from (1), but I will refrain from writing it explicitly.

Mimicking the construction of the BRST charge for Vir, Thierry-Mieg [1] was the first to define a BRST charge for $W_3$. To this effect we introduce two ghost systems $(b^{[1]}, c^{[1]})$ and $(b^{[2]}, c^{[2]})$. These are fermionic fields obeying the following operator product expansions

$$b^{[i]}(z)c^{[j]}(w) = \frac{\delta_{ij}}{z-w} + \text{reg.} \ . \quad (3)$$

The differential is then given by $d = d_0 + d_1 + d_2$ where $d_i = \oint_{C_0} \frac{dz}{2\pi i} j_i(z)$, where the currents are given by

$$
\begin{aligned}
j_0 &= c^{[1]}(T + T_{bc}^{[2]} + \tfrac{1}{2}T_{bc}^{[1]}) \\
j_1 &= c^{[2]}W \\
j_2 &= -\tfrac{8}{261}\partial c^{[2]}c^{[2]}b^{[1]}T - \tfrac{25}{348}\partial c^{[2]}c^{[2]}\partial^2 b^{[1]} - \tfrac{125}{1044}\partial^2 c^{[2]}c^{[2]}\partial b^{[1]} \\
&\quad - \tfrac{125}{1566}\partial^3 c^{[2]}c^{[2]}b^{[1]} \ ,
\end{aligned}
\quad (4)
$$

where we have introduced the Virasoro generators for the ghosts

$$T_{bc}^{[1]} = -2b^{[1]}\partial c^{[1]} - \partial b^{[1]}c^{[1]} \quad (5)$$

and

$$T_{bc}^{[2]} = -3b^{[2]}\partial c^{[2]} - 2\partial b^{[2]}c^{[2]} \ , \quad (6)$$

which generate Virasoro algebras with $c = -26$ and $c = -74$ respectively. A computation shows that $d^2 = 0$ if and only if $c_m = 100$. Thus if $\mathfrak{M}$ is any representation of $W_3$ with central charge equal to 100, we can define a complex. We shall call these representations admissible. The resulting complex is graded by ghost number, where $b^{[i]}$ (resp. $c^{[i]}$) has ghost number $-1$ (resp. 1). It is clear from the definition of the differential that $d$ has ghost number 1.

The complex can be described more explicitly if we introduce the ghost modes: $b^{[1]}(z) = \sum_n b_n^{[1]} z^{-n-2}$, $c^{[1]}(z) = \sum_n c_n^{[1]} z^{-n+1}$, $b^{[2]}(z) = \sum_n b_n^{[2]} z^{-n-3}$, and $c^{[2]}(z) = \sum_n c_n^{[2]} z^{-n+2}$. The modes obey an infinite dimensional Clifford algebra and the representation of interest is obtained as follows. Define the projectively invariant vacuum $|0\rangle$ by

$$\begin{aligned} c_n^{[1]}|0\rangle = 0 \text{ for } n \geq 2 \quad & b_n^{[1]}|0\rangle = 0 \text{ for } n \geq -1 \, , \\ c_n^{[2]}|0\rangle = 0 \text{ for } n \geq 3 \quad & b_n^{[2]}|0\rangle = 0 \text{ for } n \geq -2 \, . \end{aligned} \quad (7)$$

Then the representation space of the ghosts $\bigwedge$ is generated by $|0\rangle$ via the action of the remaining modes. It carries a representation of Vir with $c = -100$ but as far as I know it has not been shown to carry a representation of $W_3$. We now let $\mathfrak{M}$ be an admissible representation of $W_3$. Then $C \equiv C(\mathfrak{M}) = \bigwedge \otimes \mathfrak{M}$ inherits a grading by ghost number by declaring the ghost vacuum $|0\rangle$ to have zero ghost number. We let $\bigwedge^n$ denote the subspace of $\bigwedge$ of ghost number $n$ and $C^n = \bigwedge^n \otimes \mathfrak{M}$. Then $d : C^n \to C^{n+1}$ making $(C^\cdot, d)$ into a graded differential complex. Its cohomology is the BRST cohomology of $W_3$ with values in the representation $\mathfrak{M}$.

Quantum Drinfel'd-Sokolov Reduction

The $W_3$ algebra can be understood as the cohomology algebra of a given complex built out of $\widehat{sl}_3$. We start by setting up the conventions for $\widehat{sl}_3$. We first choose a basis for $sl_3$: $\{h^\pm, J_i^\pm\}$ for $i = 1, 2, 3$. The nontrivial brackets are given by

$$\begin{aligned} &[h^+, J_1^\pm] = \pm J_1^\pm & &[h^-, J_1^\pm] = \pm\sqrt{3} J_1^\pm \\ &[h^+, J_2^\pm] = \pm J_2^\pm & &[h^-, J_2^\pm] = \mp\sqrt{3} J_2^\pm \\ &[h^+, J_3^\pm] = \pm 2 J_3^\pm & &[h^-, J_3^\pm] = 0 \\ &[J_1^+, J_1^-] = \tfrac{1}{2}(h^+ + \sqrt{3} h^-) & &[J_1^\pm, J_2^\pm] = \pm J_3^\pm \\ &[J_2^+, J_2^-] = \tfrac{1}{2}(h^+ - \sqrt{3} h^-) & &[J_1^\pm, J_2^\mp] = \mp J_3^\mp \\ &[J_3^+, J_3^-] = h^+ & &[J_2^\pm, J_3^\mp] = \pm J_1^\mp \end{aligned} \quad (8)$$

and we fix an invariant inner product $(,)$ by $(h^\pm, h^\pm) = 2$ and $(J_i^+, J_j^-) = \delta_{ij}$. The affine Lie algebra $\widehat{sl}_3$ is generated by the modes of the currents $X(z)$ for $X \in sl_3$. The algebra of modes is encapsulated in the OPE:

$$X(z)Y(w) = \frac{k(X,Y)}{(z-w)^2} + \frac{[X,Y](w)}{z-w} + \text{reg.} \, , \quad (9)$$

where $k$ is the level of the representation.

We now introduce three fermionic ghost systems $(\psi_i^*, \psi^i)$, for $i = 1, 2, 3$ with OPEs

$$\psi_i^*(z)\psi^j(w) = \frac{\delta_i^j}{z-w} + \text{reg.} \quad (10)$$

and define $d_{\text{DS}} = \oint_{C_0} \frac{dz}{2\pi i} j_{\text{DS}}(z)$, where $j_{\text{DS}} \equiv -\psi^1 - \psi^2 + \sum_i \psi^i J_i^+ - \psi_3^* \psi^1 \psi^2$. This is the BRST current associated to the constraints $J_1^+ = J_2^+ = 1$ and $J_3^+ = 0$. These constraints are first-class and the BRST charge obeys $d_{\text{DS}}^2 = 0$. On the field algebra $\mathcal{A}$ generated by the $\widehat{sl}_3$ currents and the ghost systems via the operations of normal-ordered product and taking derivatives, we define an action of $d_{\text{DS}}$ via the graded commutator. This operation squares to zero since $d_{\text{DS}}$ does. Moreover, since $[d_{\text{DS}}, \cdot]$ is a derivation over the operator product and commutes with the derivative, the cohomology inherits the structure of an operator product algebra. Notice that $\mathcal{A}$ is graded by ghost number, where as usual $\psi^i$ has ghost number 1 and $\psi_i^*$ has ghost number $-1$. Under this grading the differential has ghost number one and the complex is graded. The cohomology of this complex was first studied in [**17**] where it was shown to contain the $W_3$ algebra. More recently, however, it was proven in [**15**] that the cohomology contains nothing else. Furthermore, in this paper there is an algorithm to construct generators of the algebra which close not just in cohomology (as was found in [**17**]) but actually at the level of cocycles.

To describe these generators it is convenient to introduce the following "dressed" currents:

$$\begin{aligned} &\widetilde{h}^+ = h^+ + \psi_1^*\psi^1 + \psi_2^*\psi^2 + 2\psi_3^*\psi^3 & &\widetilde{h}^- = h^- + \sqrt{3}\psi_1^*\psi^1 - \sqrt{3}\psi_2^*\psi^2 \\ &\widetilde{J}_1^+ = J_1^+ + \psi_3^*\psi^2 & &\widetilde{J}_2^+ = J_2^+ - \psi_3^*\psi^1 & &\widetilde{J}_3^+ = J_3^+ \\ &\widetilde{J}_1^- = J_1^- + \psi_2^*\psi^3 & &\widetilde{J}_2^- = J_2^- - \psi_1^*\psi^3 & &\widetilde{J}_3^- = J_3^- \end{aligned} \quad (11)$$

Then the cohomology of the complex $(\mathcal{A}^\cdot, [d_{\text{DS}}, \cdot])$ is freely generated by the cocycles

$$\begin{aligned} T = &\frac{1}{k+3}\left[\widetilde{J}_1^- + \widetilde{J}_2^- + \tfrac{1}{4}(\widetilde{h}^+)^2 + \tfrac{1}{4}(\widetilde{h}^-)^2 + (k+2)\partial \widetilde{h}^+\right] \\ W = &\mu \left[\widetilde{J}_3^- + \tfrac{1}{2}\widetilde{h}^+(\widetilde{J}_1^- - \widetilde{J}_2^-) + \tfrac{1}{2}(k+2)(\partial \widetilde{J}_1^- - \partial \widetilde{J}_2^-) \right. \\ &- \tfrac{\sqrt{3}}{6}\widetilde{h}^-(\widetilde{J}_1^- + \widetilde{J}_2^-) - \tfrac{\sqrt{3}}{36}(\widetilde{h}^-)^3 + \tfrac{\sqrt{3}}{12}(\widetilde{h}^+)^2\widetilde{h}^- \\ &\left. + (k+2)\tfrac{\sqrt{3}}{4}\widetilde{h}^+\partial \widetilde{h}^- + (k+2)\tfrac{\sqrt{3}}{12}\widetilde{h}^-\partial \widetilde{h}^+ + (k+2)^2\tfrac{\sqrt{3}}{6}\partial^2 \widetilde{h}^-\right] \, , \end{aligned} \quad (12)$$

where the constant of proportionality $\mu$ satisfies

$$\mu^2 = \frac{48}{(k+3)^3(5c_m + 22)} \, . \quad (13)$$

A computation shows that $T$ and $W$ obey (1) with central charge

$$c_m = 50 - 24(k+3) - \frac{24}{k+3} \ . \tag{14}$$

Notice that they generate an honest $\mathsf{W}_3$ algebra, not just in cohomology.[1]

The above construction extends from the algebras to their representations; giving us a way to assign to a representation (in the category $\mathcal{O}$) of $\widehat{sl}_3$ at level $k$, a representation of $\mathsf{W}_3$ with central charge $c_m$ given by (14). For example, it was shown in [17] that if $\mathfrak{N}$ is the standard (Wakimoto) free-field realization, then $H^0_{\mathrm{DS}}(\mathfrak{N})$ is precisely the two-scalar realization of $\mathsf{W}_3$ of Fateev and Zamolodchikov [10].

The representation of $\mathsf{W}_3$ can be described more explicitly as follows. Let $\bigwedge_{\mathrm{DS}}$ denote the representation space of the ghost systems $(\psi_i^*, \psi^i)$. To define it, we expand the ghosts in modes $\psi^i(z) = \sum_n \psi_n^i z^{-n-s_i}$ and $\psi_i^*(z) = \sum_n \psi_{i,n}^* z^{-n+s_i-1}$, where $s_1 = s_2 = 1$ and $s_3 = 2$. We define the projectively invariant vacuum $|0\rangle_{\mathrm{DS}}$ by

$$\psi_n^i |0\rangle_{\mathrm{DS}} = 0 \text{ for } n > -s_i \qquad \psi_{i,n}^* |0\rangle_{\mathrm{DS}} = 0 \text{ for } n \geq s_i \ ; \tag{15}$$

and we define $\bigwedge_{\mathrm{DS}}$ as the space generated by the action of the remaining modes on the vacuum $|0\rangle_{\mathrm{DS}}$. If we again declare the vacuum to have ghost number zero, $\bigwedge_{\mathrm{DS}}$ becomes graded by ghost number. We will let $\bigwedge_{\mathrm{DS}}^n$ denote the subspace of ghost number $n$. Now let $\mathfrak{N}$ be a representation of $\widehat{sl}_3$, and define a complex $C_{\mathrm{DS}} \equiv C_{\mathrm{DS}}(\mathfrak{N}) = \bigwedge_{\mathrm{DS}} \otimes \mathfrak{N}$ with differential $d_{\mathrm{DS}}$. The complex is graded by ghost number $C_{\mathrm{DS}}^n = \bigwedge_{\mathrm{DS}}^n \otimes \mathfrak{N}$ and $d_{\mathrm{DS}} : C_{\mathrm{DS}}^n \to C_{\mathrm{DS}}^{n+1}$ has ghost number one. We denote its cohomology by $H^{\bullet}_{\mathrm{DS}}(\mathfrak{N})$. It follows from the results of [15] that $H^{n \neq 0}_{\mathrm{DS}}(\mathfrak{N}) = 0$, and that $H^0_{\mathrm{DS}}(\mathfrak{N})$ is a representation of $\mathsf{W}_3$ whose central charge is given by (14). In particular, if we wish to build a $\mathsf{W}_3$-string theory we need $c_m = 100$, which forces $k$ to take the values $-\frac{15}{4}$ or $-\frac{13}{3}$.

---

[1] It is perhaps worth remarking—albeit parenthetically—that the form of $T$ and $W$ is not the canonical one, that is, they are not deformations of the quadratic and cubic casimirs in the dressed currents. For our purposes this will do, but it would still be interesting to know if there exists a canonical form for $T$ and $W$ which still close at the level of cocycles.

### New $\mathsf{W}_3$ Strings

We have just learned that any representation of $\widehat{sl}_3$ with $k = -\frac{15}{4}$ or $k = -\frac{13}{3}$, allows us to define a $\mathsf{W}_3$-string theory: we simply compute the cohomology of $d_{\mathrm{DS}}$ in that representation and then compute the cohomology of the $d$ given by (4) in the resulting space. However this is in general impracticable because the cohomology of $d_{\mathrm{DS}}$ may not be freely generated, so that we cannot simply say that we have a set of fields which generate it and then compute the $\mathsf{W}_3$ BRST cohomology in that field theory. It would be much nicer if we could work directly with the representation of $\widehat{sl}_3$ without first having to compute its $d_{\mathrm{DS}}$-cohomology. In other words, can we cook up a BRST differential defined purely in terms of $\widehat{sl}_3$ currents (and ghosts, of course) which computes the spectrum of the $\mathsf{W}_3$ string? The answer is positive as we now show. The idea is to lift the $\mathsf{W}_3$ BRST complex via $d_{\mathrm{DS}}$ to a double complex.

Let $\mathfrak{N}$ be a representation of $\widehat{sl}_3$ with $k = -\frac{15}{4}$ or $k = -\frac{13}{3}$, and let us define the total complex $C_{\mathrm{tot}} \equiv C_{\mathrm{tot}}(\mathfrak{N}) = \bigwedge \otimes \bigwedge_{\mathrm{DS}} \otimes \mathfrak{N}$. It is bigraded by taking into account the ghost numbers of $\bigwedge$ and $\bigwedge_{\mathrm{DS}}$ separately:

$$C_{\mathrm{tot}}^{p,q} = \bigwedge^p \otimes \bigwedge_{\mathrm{DS}}^q \otimes \mathfrak{N} \ . \tag{16}$$

Acting on this complex we have two differentials. On the one hand there is $D_0 : C_{\mathrm{tot}}^{p,q} \to C_{\mathrm{tot}}^{p,q+1}$ which comes induced by $d_{\mathrm{DS}}$ and $D_1 : C_{\mathrm{tot}}^{p,q} \to C_{\mathrm{tot}}^{p+1,q}$ which comes induced from the $\mathsf{W}_3$-differential (4) but where we have substituted for $T$ and $W$ the expressions (12). Because these $T$ and $W$ obey the $\mathsf{W}_3$ algebra with $c_m = 100$, $D_1^2 = 0$. Moreover, since $T$ and $W$ are $d_{\mathrm{DS}}$-cocycles, we find that $D_0$ and $D_1$ (anti)commute. In other words, the total differential $D = D_0 + D_1$ obeys $D^2 = 0$. Notice that $D$ is not homogeneous under the bigrading (16) since it has a piece ($D_0$) of bidegree $(0, 1)$ and one piece ($D_1$) of bidegree $(1, 0)$. But it is homogeneous of degree 1 relative to the total ghost number defined by

$$C_{\mathrm{tot}}^n = \bigoplus_{p+q=n} C_{\mathrm{tot}}^{p,q} \ . \tag{17}$$

Therefore $D : C_{\mathrm{tot}}^n \to C_{\mathrm{tot}}^{n+1}$. We let $H^{\bullet}_{\mathrm{tot}}$ denote its cohomology.

We claim that $H^{\bullet}_{\mathrm{tot}}$ is precisely the cohomology of the $\mathsf{W}_3$ BRST operator on the representation $H^0_{\mathrm{DS}}(\mathfrak{N})$. The proof is actually very simple. Notice that $H^{n \neq 0}_{\mathrm{DS}}(\mathfrak{N}) = 0$, hence we are almost in the position of Proposition 12.1 in [19]. In fact, this result holds for complexes which are bounded in the sense that for fixed $n$ the direct sum in (17) is finite. This is not a priori the case in our complex, but we can use the following standard trick. We notice that we can grade the complex by conformal weight. Since $D$ has conformal weight zero, it preserves this grading. So we can compute the cohomology of each graded subspace and then collate the results at the end of the day. The advantage of doing this is that for a fixed conformal weight and for a fixed total ghost number $n$, the direct sum in (17) is indeed finite. Now we can use Proposition 12.1 in [**19**] and conclude that the cohomology of $D$ is isomorphic to the cohomology of $D_1$ on the cohomology of $D_0$. But the cohomology of $D_0$ is an admissible representation of $\mathsf{W}_3$ and in that space $D_1$ is precisely the BRST operator for $\mathsf{W}_3$. The claim thus follows.

Therefore the problem now becomes to construct suitable realizations of $\widehat{sl}_3$. As mentioned before one example of this construction is the two-scalar $\mathsf{W}_3$-string, which one obtains after considering the Wakimoto realization for $\widehat{sl}_3$. It would be very interesting to know of other examples and to see if one can manage a more interesting $\mathsf{W}_3$-string spectrum.

We should remark that the only reason we get a double complex is that $T$ and $W$ defined by (12) do obey $\mathsf{W}_3$ and not up to $d_{\mathrm{DS}}$-coboundaries. In the more general case, where these coboundaries are present, the operator $D_1$ will fail to square to zero. Nevertheless it follows from homological perturbation theory that one can still concoct a total differential $D$. We sketch the proof. The thing to notice is that because $D_1^2$ does square to zero in cohomology, it follows that $D_1^2$ is not simply a $D_0$-cocycle but also a $D_0$ coboundary. In other words, that there exists an operator $D_2$ of bidegree $(2, -1)$ such that $D_1^2 + [D_0, D_2] = 0$. We let $D^{(2)} = D_0 + D_1 + D_2$. This operator $D^{(2)}$ still may not square to zero, since $[D_1, D_2] \neq 0$ necessarily. Nevertheless, one sees that $[D_1, D_2]$ is a $D_0$-cocycle of bidegree $(3, -1)$ and hence also a $D_0$-coboundary because $D_0$ has no cohomology in that bidegree. Therefore there is a $D_3$ of bidegree $(3, -2)$ such that $[D_1, D_2] + [D_0, D_3] = 0$. Continuing in this fashion, we find that one can construct a differential $D = D_0 + D_1 + D_2 + D_3 + \cdots$ such that $D^2 = 0$. Moreover $D$ is only a finite sum of $D_n$'s, since $D_n$ has bidegree $(n, 1 - n)$ and they all have conformal weight zero. Since the complex for a fixed conformal weight is bounded, we will eventually find a $D_n$ that maps everything to zero, and hence has to be zero. Furthermore, the cohomology of this $D$ still coincides with the physical spectrum of a $\mathsf{W}_3$-string; although this now follows from the computation of the spectral sequence associated to the filtration of the complex implicit in the decomposition $D = \sum_{n \geq 0} D_n$. In other words, had we taken the expression for $T$ and $W$ found in [**17**], we would have been forced to consider a filtered complex. The nice property of the $T$ and $W$ in (12) is that the filtered complex simplifies into a double complex; thus giving us two spectral sequences (as opposed to only one) with which to approximate the cohomology.

Some immediate generalizations and closing remarks

These results admit the following obvious generalization. Given any W-algebra obtained via (generalized) quantum Drinfel'd-Sokolov reduction for which a BRST theory can be constructed, one can lift its BRST operator via the Drinfel'd-Sokolov complex to either a double complex or a filtered complex (depending on whether the W-algebra closes as cocycles or only in cohomology) defined purely in terms of the affine currents. Moreover this complex computes the same cohomology.

It seems plausible that this is is not simply a way to obtain new W-string theories, but that in fact *all* W-string theories can be obtained this way. If the W-algebra is defined via the (generalized) Drinfel'd-Sokolov reduction, it seem plausible that all admissible representations can be obtained by reducing representations of the relevant affine Lie algebra. In fact, it was proposed in [**8**] based on known results for the cohomology of the Virasoro algebra, that one can compute the BRST cohomology of a W algebra by computing the BRST cohomology associated of the affine Lie algebra with values in an appropriate representation. The heuristic argument is simple. On the one hand, the W-algebra is what remains from the affine algebra when one puts as constraints a particular nilpotent subalgebra. Similarly, the BRST cohomology of the W-algebra can be thought of as the physical states of a theory where one has as constraints the W-algebra itself. Therefore, one should be able to do this all at once and simply put as constraints all of the affine Lie algebra. The physical states would now correspond to the BRST cohomology of the affine Lie algebra computed from the standard Feigin complex. This is certainly true for the Virasoro algebra (see [**8**] and references therein), but to the best of my knowledge it has not been substantiated further. Should this be true, then it could have as an interesting by-product an existence proof for the BRST operators for the W-algebras.

Let us conclude the paper with some remarks. There is at least another way from which to construct new $\mathsf{W}_3$-strings from affine Lie algebras; namely the coset construction of $\mathsf{W}_3$ from $\widehat{sl}_3 \times \widehat{sl}_3/\widehat{sl}_3$ [**18**]. This construction is not very economical, in that the full symmetry of the coset theory is larger than just $\mathsf{W}_3$; but it may lead to interesting $\mathsf{W}_3$-strings. On the other hand, representations of $W_3$ from affine Lie algebras were considered also in [**20**]; but these are Romans realizations in disguise and hence not very interesting from the point of view of $\mathsf{W}_3$-strings.